\documentclass[aps,prd,amsmath,amssymb,reprint,superscriptaddress,nofootinbib]{revtex4-2}

\usepackage[colorlinks=true,allcolors=blue]{hyperref}
\usepackage{graphicx}
\usepackage{gensymb}
\usepackage{dcolumn}
\usepackage{bm}
\usepackage[mathlines]{lineno}
\usepackage[shortlabels]{enumitem}
\usepackage{lipsum}
\usepackage{xspace}
\usepackage[nolist]{acronym}
\usepackage{color}
\usepackage[caption=false]{subfig}
\usepackage{bm}
\usepackage[normalem]{ulem}
\usepackage{verbatim}
\usepackage{xcolor}
\usepackage{booktabs}

\graphicspath{{images/}}

\newcommand\ifnotSI[1]{}

\usepackage{etoolbox}
\makeatletter
\patchcmd\linenumberpar{\@LN@parpgbrk}{\penalty\@LN@parpgpen\relax}{}{}
\makeatother

\makeatletter
\newcommand{\customlabel}[2]{%
   \protected@write \@auxout {}{\string \newlabel {#1}{{#2}{\thepage}{#2}{#1}{}} }%
   \hypertarget{#1}{#2}
}
\makeatother

\begin{document}


\title{DMRadio-Core: A new approach for GUT-scale axion searches}

\author{V.~Ankel}
\affiliation{Department of Physics, Stanford University, Stanford, CA 94305}
\affiliation{Kavli Institute for Particle Astrophysics and Cosmology, Stanford University, Stanford, CA 94305}

\author{C.~Bartram}
\affiliation{SLAC National Accelerator Laboratory, Menlo Park, CA 94025}
\affiliation{Kavli Institute for Particle Astrophysics and Cosmology, Stanford University, Stanford, CA 94305}

\author{J.~Begin}
\affiliation{Department of Physics, Princeton University, Princeton, NJ 08544}

\author{C.~Bell}
\affiliation{Department of Physics, Stanford University, Stanford, CA 94305}
\affiliation{Kavli Institute for Particle Astrophysics and Cosmology, Stanford University, Stanford, CA 94305}

\author{S.~Chaudhuri}
\affiliation{Department of Physics, Princeton University, Princeton, NJ 08544}

\author{H.-M.~Cho}
\affiliation{SLAC National Accelerator Laboratory, Menlo Park, CA 94025}
\affiliation{Kavli Institute for Particle Astrophysics and Cosmology, Stanford University, Stanford, CA 94305}

\author{J.~Corbin}
\affiliation{Department of Physics, Stanford University, Stanford, CA 94305}
\affiliation{Kavli Institute for Particle Astrophysics and Cosmology, Stanford University, Stanford, CA 94305}

\author{W.~Craddock}
\affiliation{SLAC National Accelerator Laboratory, Menlo Park, CA 94025}

\author{S.~Cuadra}
\affiliation{Laboratory for Nuclear Science, Massachusetts Institute of Technology, Cambridge, MA 02139}

\author{A.~Droster}
\affiliation{Department of Physics, Stanford University, Stanford, CA 94305}
\affiliation{Kavli Institute for Particle Astrophysics and Cosmology, Stanford University, Stanford, CA 94305}
\affiliation{SLAC National Accelerator Laboratory, Menlo Park, CA 94025}

\author{J.~Echevers}
\affiliation{Department of Nuclear Engineering, University of California, Berkeley, Berkeley, CA 94720}

\author{E.~Engelhardt}
\affiliation{Physics Division, Lawrence Berkeley National Laboratory, Berkeley, CA 94720}

\author{J.~T.~Fry}
\affiliation{Laboratory for Nuclear Science, Massachusetts Institute of Technology, Cambridge, MA 02139}

\author{K.~D.~Irwin}
\affiliation{Department of Physics, Stanford University, Stanford, CA 94305}
\affiliation{Kavli Institute for Particle Astrophysics and Cosmology, Stanford University, Stanford, CA 94305}
\affiliation{SLAC National Accelerator Laboratory, Menlo Park, CA 94025}

\author{A.~Keller}
\affiliation{Department of Physics, Stanford University, Stanford, CA 94305}
\affiliation{Kavli Institute for Particle Astrophysics and Cosmology, Stanford University, Stanford, CA 94305}

\author{R.~Kolevatov}
\affiliation{Department of Physics, Princeton University, Princeton, NJ 08544}

\author{A.~Kunder}
\affiliation{Department of Physics, Stanford University, Stanford, CA 94305}
\affiliation{Kavli Institute for Particle Astrophysics and Cosmology, Stanford University, Stanford, CA 94305}

\author{N.~Kurita}
\affiliation{SLAC National Accelerator Laboratory, Menlo Park, CA 94025}

\author{N.~Otto}
\affiliation{Department of Physics, Princeton University, Princeton, NJ 08544}

\author{E.~Pariset}
\affiliation{Laboratory for Nuclear Science, Massachusetts Institute of Technology, Cambridge, MA 02139}

\author{S.~Puranam}
\affiliation{Department of Physics, University of California, Berkeley, Berkeley, CA 94720}

\author{P.~Quassolo}
\affiliation{Applied Science and Technology, College of Engineering, University of California, Berkeley, Berkeley, CA 94720}
\affiliation{Accelerator Technology and Applied Physics Division, Lawrence Berkeley National Laboratory, Berkeley, CA 94720}

\author{N.~M.~Rapidis}
\email{rapidis@stanford.edu}
\affiliation{Department of Physics, Stanford University, Stanford, CA 94305}
\affiliation{Kavli Institute for Particle Astrophysics and Cosmology, Stanford University, Stanford, CA 94305}

\author{C.~P.~Salemi}
\affiliation{Department of Physics, University of California, Berkeley, Berkeley, CA 94720}
\affiliation{Physics Division, Lawrence Berkeley National Laboratory, Berkeley, CA 94720}

\author{M.~Simanovskaia}
\affiliation{Department of Physics, Stanford University, Stanford, CA 94305}
\affiliation{Kavli Institute for Particle Astrophysics and Cosmology, Stanford University, Stanford, CA 94305}

\author{J.~Singh}
\affiliation{Department of Physics, Stanford University, Stanford, CA 94305}
\affiliation{Kavli Institute for Particle Astrophysics and Cosmology, Stanford University, Stanford, CA 94305}

\author{P.~Stark}
\affiliation{Department of Physics, Stanford University, Stanford, CA 94305}
\affiliation{SLAC National Accelerator Laboratory, Menlo Park, CA 94025}
\affiliation{Kavli Institute for Particle Astrophysics and Cosmology, Stanford University, Stanford, CA 94305}

\author{E.~C.~van~Assendelft}
\affiliation{Department of Physics, Stanford University, Stanford, CA 94305}
\affiliation{Kavli Institute for Particle Astrophysics and Cosmology, Stanford University, Stanford, CA 94305}

\author{K.~van~Bibber}
\affiliation{Department of Nuclear Engineering, University of California, Berkeley, Berkeley, CA 94720}
\affiliation{Physics Division, Lawrence Berkeley National Laboratory, Berkeley, CA 94720}

\author{K.~J.~Vetter}
\affiliation{Laboratory for Nuclear Science, Massachusetts Institute of Technology, Cambridge, MA 02139}

\author{K.~Wells}
\affiliation{Department of Physics, Stanford University, Stanford, CA 94305}

\author{J.~Wiedemann}
\affiliation{Department of Physics, Princeton University, Princeton, NJ 08544}

\author{L.~Winslow}
\affiliation{Laboratory for Nuclear Science, Massachusetts Institute of Technology, Cambridge, MA 02139}

\author{D.~Wright}
\affiliation{Department of Physics, Stanford University, Stanford, CA 94305}
\affiliation{Kavli Institute for Particle Astrophysics and Cosmology, Stanford University, Stanford, CA 94305}

\author{A.~K.~Yi}
\affiliation{SLAC National Accelerator Laboratory, Menlo Park, CA 94025}
\affiliation{Kavli Institute for Particle Astrophysics and Cosmology, Stanford University, Stanford, CA 94305}

\author{B.~F.~Zemenu}
\affiliation{Department of Physics, Stanford University, Stanford, CA 94305}
\affiliation{SLAC National Accelerator Laboratory, Menlo Park, CA 94025}
\affiliation{Kavli Institute for Particle Astrophysics and Cosmology, Stanford University, Stanford, CA 94305}

\date{\today}

\begin{abstract}
Searches for QCD axions with masses in the neV$/c^2$ mass range are strongly motivated by new physics at the GUT scale and by well-motivated pre-inflationary axion symmetry breaking scales. This parameter space is challenging to probe due to the small axion–photon couplings, which typically require large, high-field magnets with substantial stored energy. In this paper, we propose a new experimental geometry based on a narrow-bore, segmented solenoid that optimizes the collection of the axion-induced signal using LC resonators outside the high-field region of the magnet bore. This alternative optimization significantly reduces the required stored magnetic energy while preserving sensitivity, thus enabling a near-term experiment in the 30--200 MHz (120--830 neV$/c^2$) range, with a cost-effective, staged scaling to a GUT-scale experiment in the 100 kHz--30 MHz (\mbox{0.4--120 neV$/c^2$}) range.

\end{abstract}

\maketitle


%
%

\newcommand{\gagg}{\ensuremath{g_{a\gamma\gamma}}\xspace}
\newcommand{\gagggann}{\ensuremath{g_{a\gamma\gamma}g_{aNN}\xspace}}
\newcommand{\gagggaee}{\ensuremath{g_{a\gamma\gamma}g_{aee}\xspace}}

\newcommand{\rhoDM}{\ensuremath{\rho_{\rm DM}}\xspace}
\newcommand{\Jeff}{\ensuremath{\mathbf{J}_{\rm eff}}\xspace}

\newcommand{\cPU}{\ensuremath{c_{\rm PU}}\xspace}
\newcommand{\VPU}{\ensuremath{V_{\rm PU}}\xspace}
\newcommand{\LPU}{\ensuremath{L_{\rm PU}}\xspace}
\newcommand{\Leff}{\ensuremath{L_{\rm eff}}\xspace}

\newcommand{\DMR}{\mbox{DMRadio}\xspace}
\newcommand{\DMRp}{\mbox{\DMR-Pathfinder}\xspace}
\newcommand{\DMRL}{\mbox{\DMR-50L}\xspace}
\newcommand{\DMRm}{\mbox{\DMR-m$^3$}\xspace}
\newcommand{\DMRGUT}{\mbox{\DMR-GUT}\xspace}

\newcommand{\ABRA}{\mbox{ABRACADABRA}\xspace}
\newcommand{\ABRAten}{\mbox{ABRACADABRA-10\,cm}\xspace}


%
%

\begin{acronym}
\acro{SM}{Standard Model}
\acro{QED}{quantum electrodynamics}
\acro{QCD}{quantum chromodynamics}
\acro{BSM}{beyond the standard model}
\acro{DM}{dark matter}
\acro{CDM}{cold dark matter}
\acro{GUT}{grand unification theory}
\acro{WIMP}{weakly interacting massive particle}
\acro{SHM}{Standard Halo Model}
\acro{ppm}{part-per-million}
\acro{ppb}{part-per-billion}

\acro{ADM}{axion dark matter}
\acro{ALP}{axion-like particle}
\acro{PQ}{Peccei-Quinn}
\acro{PQWW}{Peccei-Quinn-Wilczek-Weinberg}
\acro{KSVZ}{Kim-Shifman–Vainshtein–Zakharov}
\acro{DFSZ}{Dine–Fischler–Srednicki–Zhitnitsky}
\acro{LSW}{light shining through wall}

\acro{DP}{dark photon}

\acro{DR}{dilution refrigerator}
\acro{PT}{pulse tube}
\acro{OFHC}{oxygen-free, high-conductivity}

\acro{TE}{transverse electric}
\acro{TM}{transverse magnetic}
\acro{TEM}{transverse electromagnetic}
\acro{MQS}{magneto-quasistatic}
\acro{SQL}{standard quantum limit}
\acro{QND}{quantum non-demolition}

\acro{DFT}{discrete Fourier transform}
\acro{FFT}{fast Fourier transform}
\acro{SNR}{signal-to-noise ratio}
\acro{PSD}{power spectral density}

\end{acronym}

%
\acrodefplural{PSD}{power spectral densities}
\acrodefplural{GUT}{grand unification theories}
\acrodefplural{ppm}{parts-per-million}
\acrodefplural{ppb}{parts-per-billion}


\section{Introduction}
\label{sec:Intro}

The particle nature of dark matter remains one of the central open questions in modern particle physics. Originally proposed as a solution to the strong CP problem, the quantum chromodynamics (QCD) axion \cite{Peccei:1977hh,Peccei:1977ur,Weinberg1977,Wilczek:1977pj} is a well-motivated dark matter candidate \cite{Preskill1983,Abbott:1982af,Dine1983}. While the allowed mass range spans many orders of magnitude, higher energy theories such as string theories \cite{Green:1984sg,Svrcek:2006yi,Conlon:2006tq,Acharya:2010zx,Cicoli:2012sz,Halverson:2019cmy,Witten:1984dg,Benabou:2025kgx} and grand unified frameworks \cite{Wise:1981ry,Ballesteros:2016xej,Co2021rt,Ernst:2018bib,DiLuzio:2018gqe,Ernst:2018rod,FileviezPerez:2019fku,FileviezPerez:2019ssf,Agrawal:2022lsp} point to axions with masses in the neV/$c^2$ range.

Axion dark matter can be detected through its coupling to Standard Model photons. In the neV/$c^2$ mass range, the axion Compton wavelength is $\gtrsim$ 1 m, motivating the use of lumped-element LC resonators to resonantly enhance the signal induced in the presence of large magnetic fields \cite{Cabrera2008,sikivie2014proposal,PhysRevLett.117.141801,PhysRevD.92.075012}. This approach contrasts with searches in the $\mu$eV/$c^2$ range, which enhance the signal through resonant microwave cavity modes in strong magnetic fields. Key aspects of the lumped-element technique have been explored in a series of demonstrator-scale experiments, including ABRACADABRA-10 cm \cite{PhysRevLett.127.081801,PhysRevLett.122.121802}, ADMX-SLIC \cite{Crisosto:2019fcj}, SHAFT \cite{Gramolin2020a}, and the hidden-photon search DMRadio-Pathfinder \cite{10.1007/978-3-030-43761-9_16}.

A central challenge lies in scaling this technique to larger and, hence, more sensitive detectors, particularly given that the required large magnetic volumes dominate the cost of these experiments. In this paper, we propose a new experimental concept, DMRadio-Core, designed as a scalable, near-term implementation that explores an alternative geometry optimization: reducing the magnet bore size while employing an extended structure that couples to the ac signal induced outside the high-field region (see Fig. \ref{fig:CoreBasics}). In conventional solenoidal geometries, the pickup is located within the magnet bore, and therefore the sensitivity is limited by the bore size and field. In the Core geometry, the pickup is situated primarily outside the bore, and thus its volume is no longer limited by the bore size. This allows for smaller magnets  in experiments that achieve comparable sensitivity to their conventional solenoidal counterparts at similar magnetic field strengths. The DMRadio-Core experiment is intended both as a physics experiment in its own right and as a pathfinder towards a future, larger-scale realization capable of probing GUT-scale axion parameter space. Within the broader DMRadio program, complementary geometries (Fig. \ref{fig:Comparison}) are under construction with the toroidal DMRadio-50L experiment \cite{Ankel:2026cef,DMR50LDesign,Rapidis:2022gti,Benabou:2022qpv} and the solenoidal CAL-Pathfinder \cite{calaxion}. In addition, detailed design studies were performed for the solenoidal DMRadio-m$^3$ proposal \cite{Brouwer:2022DMRm,DMRadio:2023igr}.

This paper is organized as follows. We first introduce the principles underlying lumped-element axion detection and build toward the design of the proposed DMRadio-Core experiment, which targets operation in the 290--580 neV$/c^2$ (70--140\,MHz) range, with possible extensions to cover 30--200\,MHz. We then outline how these concepts extend to a future GUT-scale experiment with significantly larger magnetic volumes and enhanced sensitivity. In Section \ref{sec:core_foundations}, we describe the foundational concepts of the Core geometry, including the mechanism by which the axion field induces currents on the pickup and the role of boundary conditions in the lumped-element regime, which favor large detection volumes even outside regions of DC magnetic field. We also compare the Core, toroidal, and solenoidal geometries. In Section \ref{sec:geometryspecifics}, we present a concrete implementation based on axially stacked pickups, along with considerations for signal readout and phase coherence. Section \ref{sec:CoreExp} details the design of the proposed DMRadio-Core experiment, including magnet specifications, pickup configuration, and the numerically evaluated sensitivity to DFSZ axions. Finally, in Section \ref{sec:GUT}, we present an extrapolation to a future GUT-scale experiment, including representative design parameters and projected sensitivity.

\section{Foundations of the Core geometry and DMRadio}
\label{sec:core_foundations}

The primary figure of merit of resonant axion haloscopes is the scan rate: the rate at which the experiment tunes its resonance frequency such that it can achieve sensitivity to axions at a given axion-photon coupling $g_{a\gamma\gamma}$ and a given signal-to-noise ratio (SNR). For experiments operating in the deep magneto-quasistatic (MQS) limit where the Compton wavelength of the axion is much larger than the spatial extent of the pickup and the displacement current term in Maxwell's equations can be ignored, parameters of interest that set the scan rate scale as:
\begin{equation}
    \frac{d\nu_r}{dt}\propto g_{a\gamma\gamma}^4\nu_rB_0^4V_{\text{pu}}^{10/3}Q
    \label{eq:scanratescaling}
\end{equation}
where $\nu_r$ is the resonance frequency, $B_0$ is the peak strength of the dc magnetic field, $V_\text{pu}$ is the pickup volume, and $Q$ is the quality factor \cite{DMRadio:2022jfv}. As such, experiments benefit from large pickup volumes, strong magnetic fields, and high quality factors.

\begin{figure}
    \centering
    \includegraphics[width=.625\columnwidth]{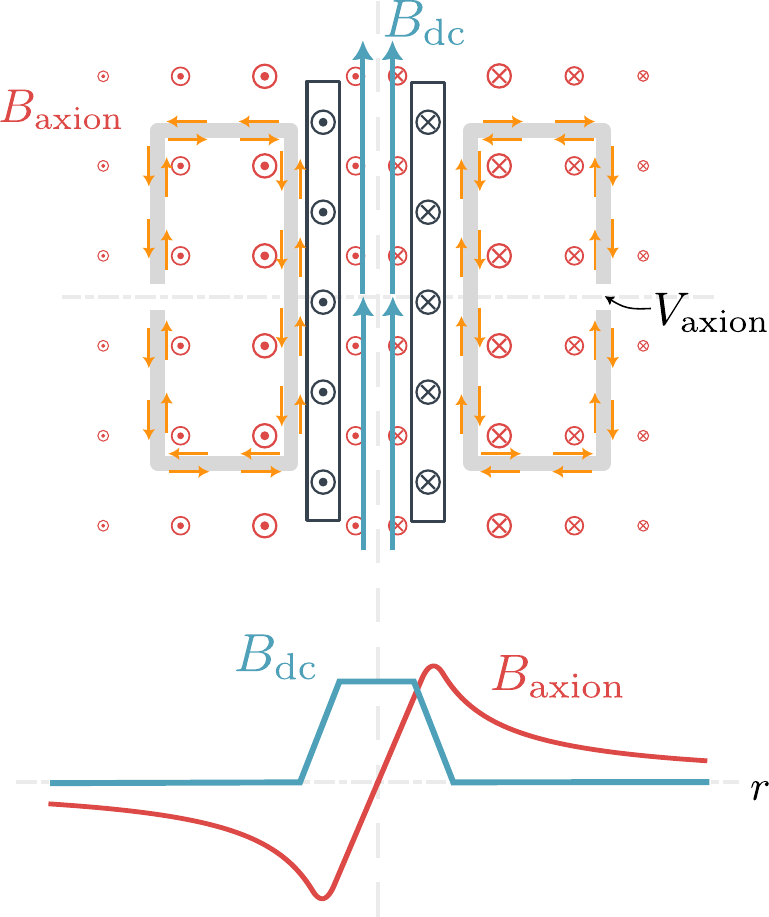}
    \caption{ Top: The fundamental concept of the Core geometry. The pickup (gray) with a slit drives physical electron currents (orange) in response to the change in the ac axion-induced flux (red) on the outside of a solenoidal magnet. A voltage develops across the slit due to the change in magnetic flux. Bottom: The dc magnetic field is uniform within the solenoid but the axion-induced field extends beyond the solenoid bore, following a $r^{-1}$ power law in the ideal case of an infinite solenoid. The $y$-scale is in arbitrary units for both curves.
    }
    \label{fig:CoreBasics}
\end{figure}

\begin{figure*}
    \centering
    \includegraphics[width=.655\textwidth]{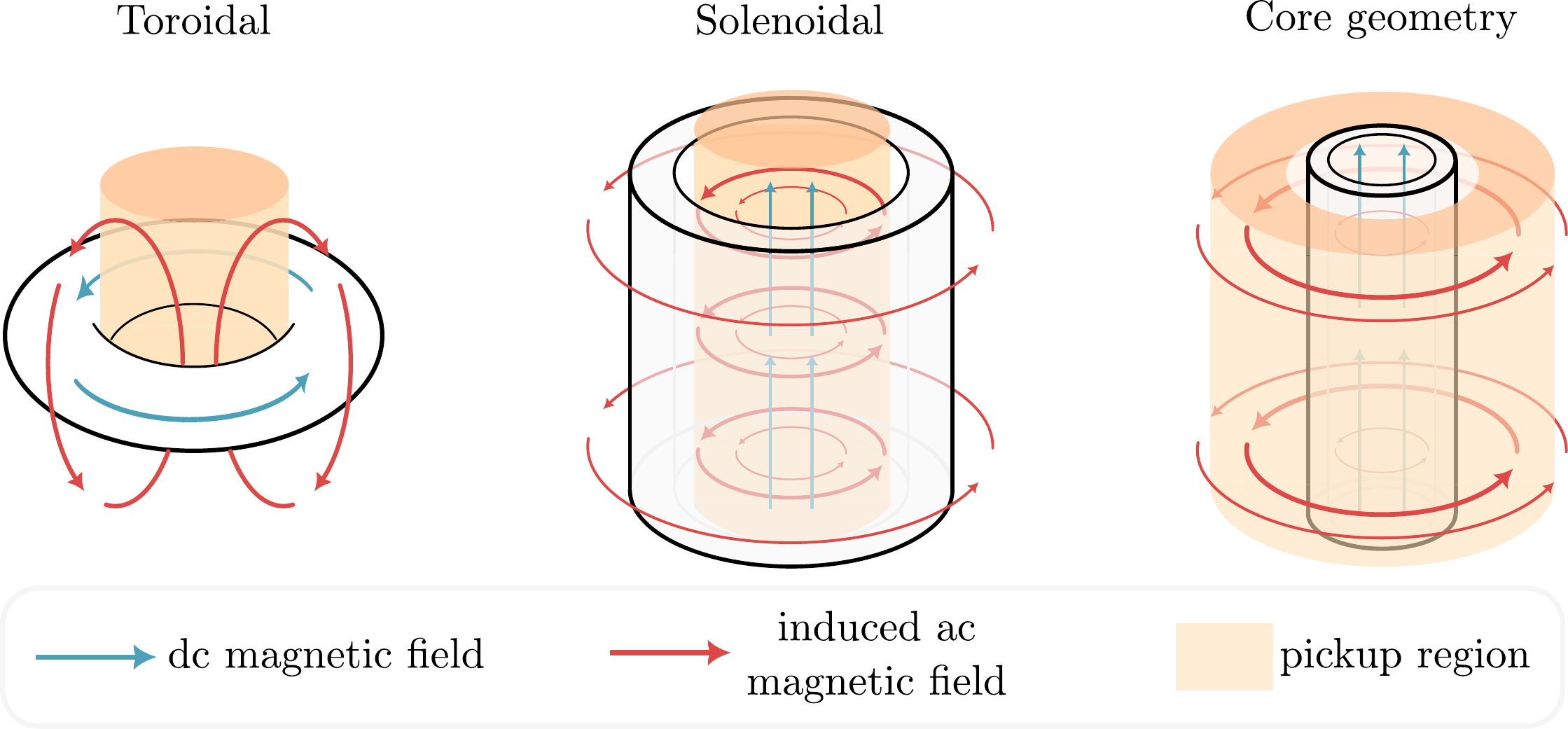}
    \caption{3D schematic of the toroidal (left), solenoidal (center), and Core (right) geometries. The solenoidal and Core geometries both use solenoidal magnets but, by coupling to the induced ac field outside of the magnet, the Core geometry can employ more compact magnets than the solenoidal counterpart.
    }
    \label{fig:Comparison}
\end{figure*}

In solenoidal geometries (e.g. DMRadio-m$^3$ and CAL-Pathfinder) and toroidal geometries (e.g. DMRadio-50L), as illustrated in Figure \ref{fig:Comparison}, the size of the pickup structure is limited by the size of the magnet. Specifically, in the case of solenoidal geometries which use a coaxial pickup within the bore of a solenoidal magnet, the outer diameter of the pickup is set by the inner diameter of the solenoidal magnet bore.  Any axial extension of the pickup beyond the height of the solenoidal magnet is generally unfavorable due to the increased inductance for a constant flux. This reduces the coupled energy $U_c=\Phi^2/2L$, where $\Phi$ is the total magnetic flux, and $L$ is the inductance \cite{DMRadio:2023igr}. Conversely, reducing the size of the pickup while keeping the magnet constant reduces the axion-induced magnetic energy that is coupled to the pickup. Similarly, in the case of toroidal geometries, which employ a toroidal magnet, the size of the pickup inductor is determined by the inner radius and height of the toroidal magnet. An inductor that matches those dimensions has the maximum sensitivity to an axion-induced signal. As magnets present the dominant cost of an experiment, it is  especially beneficial to have an experimental geometry where the magnet no longer limits the size of the pickup. Specifically, in lumped-element experiments where the wavelength of the induced signal is much longer than the detector, the external metallic shield of the experiment imposes boundary conditions on the electric and magnetic fields induced by the axion signal. These boundary conditions force the signal to be much weaker than a signal produced in empty space \cite{PhysRevD.92.075012}. If the external shield, and hence the pickup volume, can be expanded such that the boundary conditions are enforced  at larger distances, the experimental sensitivity to the dark matter signals increases.

The \textit{Core} geometry decouples the size of the pickup from the physical extent of the magnet. For a long solenoid, the dc magnetic field is uniform within the bore and diminishes to near-zero values radially outside the coils. Axion dark matter couples to the dc magnetic field, generating an ac field equivalent to that produced by an effective current density of the form:
\begin{equation}
    \mathbf{J}_\text{eff}(\textbf{x},t)=g_{a\gamma \gamma}\frac{\sqrt{\hbar c}}{\mu_0}\sqrt{2\rho_\text{DM}}\cos\left(\frac{m_ac^2}{\hbar}t\right)\mathbf{B}_\text{dc}(\textbf{x})
    \label{eq:Jeff}
\end{equation}
where $\rho_\text{DM}$ is the local dark matter density of $0.45 \text{ GeV cm}^{-3}$ \cite{de_Salas_2021} and $m_a$ is the axion mass. This current induces oscillating magnetic fields both inside and outside the solenoid as illustrated in Figure \ref{fig:CoreBasics}. Traditionally, in geometries like \DMRm and microwave cavities, the experiment is designed to pick up the electromagnetic response induced by the effective current density within the bore of the solenoid. The Core geometry removes this constraint by coupling to the induced electromagnetic response outside of the solenoid. This is a region where the dc field is near zero but where a large fraction of the axion-induced ac magnetic field resides.

The ac magnetic field drives currents on the surface of a metallic pickup with a slit interrupting the shielding current. This pickup is placed on the outside of the solenoidal magnet. In this region, and in  the case of an ideal infinite solenoid, the axion-induced ac magnetic field scales as $B_\text{axion}\sim r^{-1}$ as determined by Amp\`ere's law. The axion ultimately induces a voltage across the slit, which drives a current that can be measured using a flux-to-voltage amplifier.  A model pickup and the induced fields are shown in Figure \ref{fig:CoreBasics}. This argument might at first appear inconsistent with the cavity mode formalism where the cavity haloscope is designed to maximize the overlap integral of the axion-induced current with the profile of the cavity mode. However, this is not the case as is detailed in Appendix \ref{sec:Cavity_v_LumpedElement}.

The Core geometry has the additional advantage that the pickup can exist in a region with low dc magnetic field outside of the solenoid. By judicious choices of magnet geometries, the dc magnetic field can be smaller than the lower critical field, $H_{c1}$, of niobium, thus making the use of superconducting pickup structures natural.  Such superconducting pickups enhance the quality factor of the resonator and therefore increase the sensitivity of the experiment.

It is worth emphasizing the benefits of the Core geometry compared to the toroidal and solenoidal geometries. The Core geometry benefits from the ease of assembly of solenoidal magnets compared to toroidal magnets. In addition, the significantly reduced magnet bore size of the Core geometry compared to the solenoidal geometry presents the dominant benefit. Compact magnets drastically reduce the cost of such an experiment for similar sensitivities. The Core geometry also shares the benefit of the toroidal geometry wherein the dc magnetic fields naturally avoid the regions where superconducting elements exist. As such, the Core geometry adopts the benefits of the toroidal and solenoidal geometries while additionally reducing the size and stored energy of the dc magnets that are required.

\section{Optimization of the Core geometry}
\label{sec:geometryspecifics}

The Core geometry shown in Figure \ref{fig:CoreBasics} presents the fundamental idea behind DMRadio-Core. However, any such geometry requires shielding from external electromagnetic interference and would also otherwise operate within a cryostat. As such, further boundary conditions must be enforced in the form of an external metallic structure. The design of an optimal Core geometry must avoid structures that would support low frequency cavity modes that pose a concern for such experiments due to a decrease in the scan rate at frequencies around the cavity mode frequency \cite{DMRadio:2023igr}. However, as previously mentioned, the enforced boundary conditions reduce the field inside the pickup. With these constraints, finite element modeling (FEM) simulations with COMSOL allow for an optimization of the pickup by considering its scan rate. We reformulate the equation of the scan rate, as first presented in Equation \ref{eq:scanratescaling}, such that it is written in terms of parameters that are quantifiable through experimental measurements:
\begin{equation}
\begin{split}
\frac{d\nu}{dt}&=\frac{\pi(6.4\times10^5)}{\text{SNR}^2}\frac{\hbar^2}{16c^8m_a^4}\times\\
&\qquad\frac{|V(m_a,\mathbf{B}(\mathbf{r}),g_{a\gamma\gamma})|^4Q(\omega_r)\bar{\mathcal{G}}[\omega_r,T,\eta(\omega_r)]}{L_\text{eff}(\omega_r)^2}.
\end{split}
\label{eq:scanrate}
\end{equation} 
Here $|V(m_a,\mathbf{B}(\mathbf{r}),g_{a\gamma\gamma})|$ is the magnitude of the axion-induced voltage across the slit of the pickup that scales linearly with the magnetic field and $g_{a\gamma\gamma}$, $Q$ is the quality factor of the resonator, $\bar{\mathcal{G}}$ parametrizes noise physics as described in \cite{DMRadio:2022jfv} and \cite{Chaudhuri:2018rqn}, $L_\text{eff}$ is the effective inductance of the pickup and readout, and the numerical prefactor is determined by the physics of the dark matter halo. The FEM simulations provide numerical results for $|V|$, $Q$, and $L_\text{eff}$ for a given magnet and pickup design. Specifically, these simulations have shown that an optimized geometry involves a fully enclosed pickup that penetrates through the core of a solenoid.

\subsection{Pickup design}

\begin{figure}
    \centering
    \includegraphics[width=.70\columnwidth]{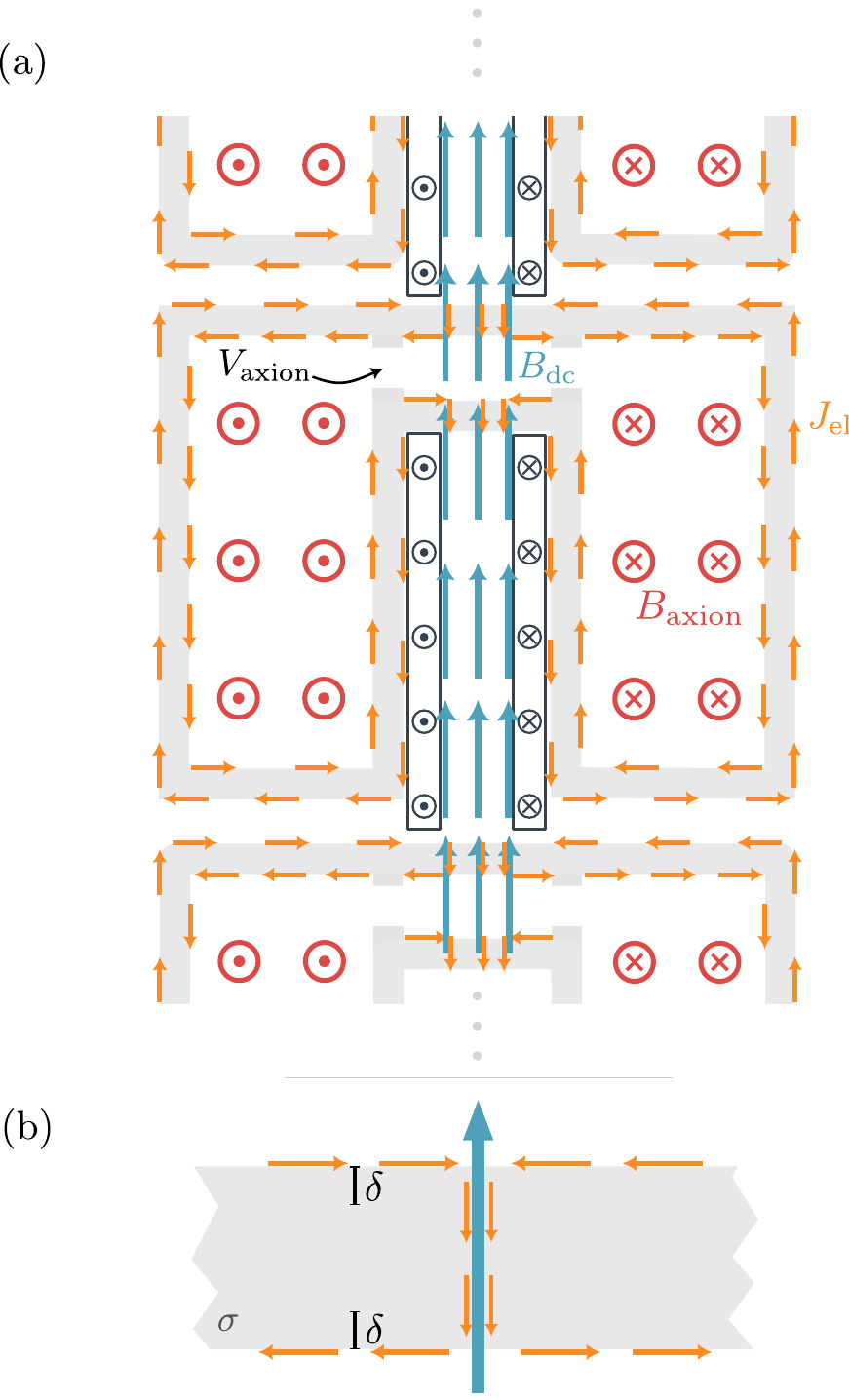}
    \caption{(a) A 2D cross section of stacked closed metallic structures. The induced current flows on the inner walls of each structure into regions of low dc field where it induces an ac magnetic field. This ac field develops a voltage across the slit (shown in figure) and is measured using a flux-to-voltage amplifier (not shown) connected to this slit. Tuning elements (not shown) are also connected across this slit. The magnitude of the current in this illustration is constant for illustrative purposes and does not take into account variations in magnitude due to current densities reducing at larger radial distances.  (b) A schematic of the effect of bulk electron shuttling in a slab of metal. The dc magnetic field line (blue) intersects a metal of conductivity $\sigma$ which has a skin depth $\delta$ at a particular frequency. A physical electron current (orange) of equal magnitude to the effective axion current is driven through the bulk of the metal, oscillating at the frequency set by the axion mass, $\nu_a=m_ac^2/h$. As such, axion dark matter and dc magnetic fields can drive currents within structures that are otherwise fully shielded from their environment by the skin depth of the metal.
    }
    \label{fig:BES}
\end{figure}

The pickup shown in Figure \ref{fig:BES}a employs an enclosed structure which intersects the magnetic field lines produced by a solenoid. The effective axion current drives physical electron currents on the walls of the pickup from the region of high magnetic field into the regions where the dc magnetic field is low. In this low field region, the physical electron current drives ac magnetic fields which induce a voltage across the slit. The voltage and impedance across this slit determine the scan rate of an experiment since the amplifiers and the tuning elements are also connected across this slit. Since the pickup is fully enclosed, only the impedance of the interior segment of the pickup determines the induced ac fields and, as a result, any metallic structures such as magnet coils do not influence the response of the structure since they are shielded by the skin depth of the pickup.

The behavior of the physical electron current illustrated with this pickup geometry relies on the effect of bulk electron shuttling (BES) as derived in Appendix D of \cite{DMRadio:2023igr}. The physical basis of BES can be understood as a specific case of screening currents in such structures. The effective axion current  flows in the direction of the dc magnetic field lines through the metal and therefore an equal and opposite screening current must flow through the bulk of the metal to satisfy Amp\`ere's law.  It is important to emphasize that this current flows within the bulk of the metal, irrespective of the skin depth of the metal. This effect is illustrated schematically in Figure \ref{fig:BES}b. When observing only the inner portion of the pickup, it may appear that there exist sources and sinks of electrons. However, these are fully explained by the BES effect where electrons move into the bulk of the material and are not created or destroyed.

While the impedance of such a structure across the slit is non-analytic and must either be measured or simulated numerically, it can be roughly modeled as an inductor (the large superconducting pickup region outside the magnet) in parallel with a capacitor (the parallel plate  portion traversing the magnetic field). A lumped element that is either capacitive or inductive connects across the slit to load the impedance of the structure and to, therefore, tune the resonance frequency. The current induced within this structure can subsequently be picked up with a flux-to-voltage amplifier such as a dc SQUID or a device operating below the standard quantum limit, such as a  radiofrequency quantum upconverter (RQU) \cite{Kuenstner:2022gyc}. The superconducting readout devices and  tuning elements (not shown in Figure \ref{fig:BES}a) exist within the pickup in the low dc field region such that they can operate below their $H_{c1}$ values.

To achieve increased sensitivity to axions, a larger structure is preferred. However, to avoid cavity modes within the target frequencies of an experiment, it is instead beneficial to segment the pickups by stacking them axially in a solenoidal magnet that is also segmented axially. This allows for the axion induced signal to be added coherently across multiple structures. The scaling of the scan rate by co-adding signals from different structures is discussed in the following subsection.

The section crossing the bore of the solenoid must be made of non-superconducting metal such as copper since the high dc magnetic fields would otherwise drive a superconductor into its normal state. While the resistive copper contributes to loss that can degrade the quality factor of the resonator, most of the current flow is across the superconducting section, therefore reducing resistive losses and protecting the quality factor.

\subsection{Scaling of scan rate with multiple pickups}

Since this geometry employs multiple pickups, the signals and noise from each pickup contribute to the total response of the experiment. Specifically, their SNRs add in quadrature as:
\begin{equation}
    \text{SNR}_\text{tot}^2=\sum_i^N\text{SNR}_i^2
\end{equation}
where there are $N$ pickups, and $\text{SNR}_i$ is the SNR from the $i$-th pickup. 

The scan rate, $d\nu/dt$, is the rate at which a resonant experiment must tune its resonance frequency while maintaining sensitivity to an axion at a given coupling strength. To calculate the total scan rate of the combined pickups, we consider the scan rate of each pickup if it were operating on its own, $(d\nu/dt)_i$. The summed scan rate over $N$ pickups when all are tuned in unison is:
\begin{equation}
   \left[\frac{d\nu}{dt}(\nu)\right]_\text{tot}=\left[\sum_i^N\sqrt{\left[\frac{d\nu}{dt}(\nu)\right]_i}\ \right]^2.
\end{equation}
This equation assumes that the signals add coherently, and would not be relevant for pickups operating at times separated by more than the axion coherence time.

Concerns related to phase delays of the electromagnetic signal often arise in axion experiments. Any relative phase delay in the case of the DMRadio-Core pickups can be calibrated and corrected for in software, thanks to the low frequencies at which the experiment operates.

Similarly, the de Broglie wavelength of the axion dark matter must be considered as well. The de Broglie wavelength defines the distance over which the axion signal is expected to be coherent, as set by dark matter halo dynamics. The minimum de Broglie wavelength probed by DMRadio-GUT is 10 km, much larger than the physical extent of the experiment. As a result, the induced signal within the experiment will be phase coherent across the extent of the experiment. By extension, the voltage induced across all the slits will also be phase coherent and will not pose a concern for the readout scheme.

\section{The \DMR-Core Experiment}
\label{sec:CoreExp}
In this section we present DMRadio-Core, an experiment utilizing the Core geometry to probe axions with DFSZ sensitivity in the 70--140 MHz frequency range corresponding to 290--580 neV/$c^2$ masses, with possible extensions to cover 30--200 MHz. Specifically, the DMRadio-Core experiment covers the originally proposed  \DMRm parameter space. As we will  describe below, DMRadio-Core utilizes a segmented $\sim$5 T peak field solenoidal magnet with a 37 cm bore diameter, 60 cm outer diameter, and 2.95 m height. This is in contrast with \DMRm, which utilizes a $\sim$6.7 T peak field solenoidal magnet, with a 1.45 m  bore diameter, 1.72 m outer diameter, and 1.35 m height. As such, the magnet for DMRadio-Core contains an order of magnitude less stored energy than the magnet of \DMRm, while allowing the two experiments to achieve comparable sensitivity to DFSZ axions. 

\subsection{Magnet and pickup geometry}

\begin{figure*}[ht]
    \centering
    \includegraphics[width=.985\textwidth]{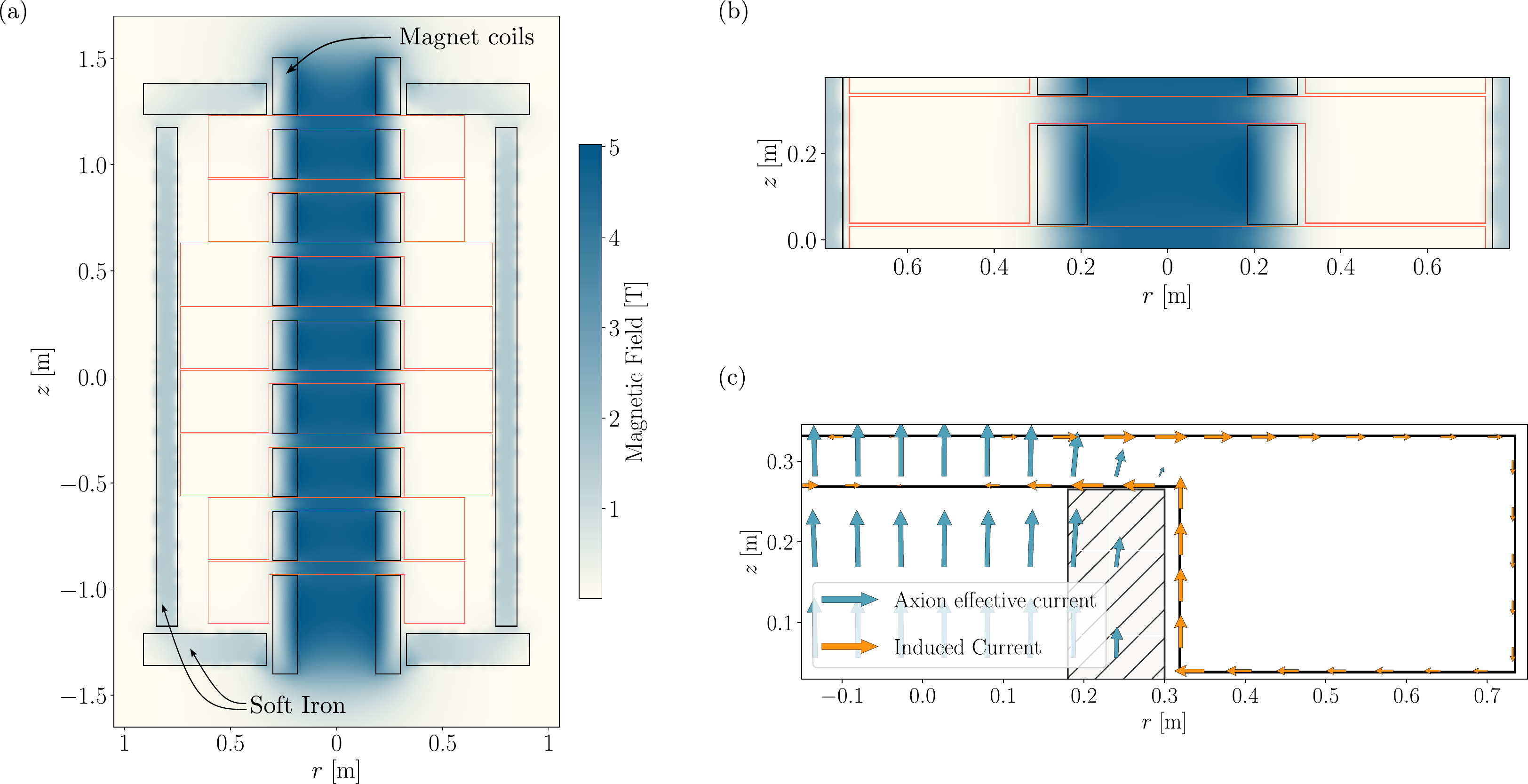}
    \caption{(a) The 2D magnetic field profile of the DMRadio-Core magnet with the nine coil segments and the soft iron cross sections overlaid. The dc magnetic field peaks at 5 T within the center of the 37 cm bore solenoid. The majority of the magnetic field lines return either through the soft iron or at larger radial positions, thus creating the low-field region where the pickup is primarily housed. The topmost two and bottommost two pickups have $d_o=120$ cm, and the central four have $d_o=146$ cm. The different impedances of these pickups enable the continuous sensitivity of DMRadio-Core across all frequencies by omitting specific pickups at frequencies where their scan rate is unfavorable. The fine spatial variation of the magnetic field in the soft iron is attributed to simulation noise and does not meaningfully affect the sensitivity of the experiment. (b) A zoomed-in cross section of a 146 cm diameter pickup (orange) overlaid on top of the dc magnetic field. The narrow sections of the pickup are placed in the high dc field region of the solenoid to source the axion signal. A large pickup area exists in the low field region outside of the magnet coils. A slit between the narrow portion and the large pickup area determines where the voltage is read out and where the readout and tuning elements (not shown) are connected to the pickup. (c) An illustration of the COMSOL simulated axion effective current (teal) which flows along the magnetic field lines, and the simulated induced physical current (orange) which flows on the inner walls of the pickup. The magnitude of the induced current grows at larger radii within the bore of the magnet and can be understood through the effect of BES as each magnetic field line contributes to additional current that will flow radially outwards within the structure.
    }
    \label{fig:DMCoreGeometry}
\end{figure*}

The DMRadio-Core  5 T peak field solenoidal magnet, shown in Figure \ref{fig:DMCoreGeometry}a, is composed of nine segments with a 37 cm bore, a 60 cm outer diameter, and a total height of 2.95 m, corresponding to a total magnetic field energy of 4.3 MJ. These segments contain 3 mm diameter NbTi wire carrying a current of 330 A. A cross section of the magnet is illustrated in Figure \ref{fig:DMCoreGeometry}a. Three segments of soft iron redirect the returning magnetic field lines such that there exists a large region where $B_\text{dc}<170$ mT, capable of housing the superconducting niobium portion of the pickup.

The pickup used in DMRadio-Core is illustrated in orange in Figures \ref{fig:DMCoreGeometry}a and b. It consists of a large superconducting niobium, or niobium plated, pickup region outside of the solenoid coils and two closely spaced copper plates traversing the region of high magnetic field. To avoid frequencies of low scan rates due to unfavorable impedances of pickups, pickups of two different outer radii are used in this experiment. Across all pickups, the copper plates have a 6 cm spacing, and the large pickup region outside of the magnet is 29 cm in height and 64 cm in inner diameter. Four pickups have an outer diameter of 120 cm (topmost two and bottommost two pickups in Figure \ref{fig:DMCoreGeometry}a) and four pickups have an outer diameter of 146 cm (central four pickups in Figure \ref{fig:DMCoreGeometry}a). 

The pickup tuning elements and readout components are connected to a slit near the interface region between the niobium pickup and the copper plates. Since the spatial extent of the tuning elements and readout components is smaller than the pickup, they are connected only across a 30$\degree$ portion of the slit. This means that the impedance is measured using a port of fractional angular width, rather than a port that covers the full azimuthal extent of the slit.

\subsection{Science reach and scan time}
\label{sec:sciencereachformalism}

The formalism and FEM techniques established in \cite{DMRadio:2023igr} based on the results of \cite{Chaudhuri:2018rqn} establish a rigorous approach to extracting the sensitivity of any single-moded resonant axion experiment. This formalism is agnostic to whether the experiment operates in the deep-MQS or cavity limits, as long as it is single-moded. Therefore, even though the sensitivity of DMRadio-Core presented here extends to 140 MHz where the Compton wavelength is approximately 2 m and the experiment is no longer in the deep-MQS limit, the techniques from \cite{DMRadio:2023igr} are appropriate to extract the scan rate of the experiment. Using FEM models in COMSOL, we extract the voltage induced by a DFSZ axion across the slit of this structure. The same simulations provide the impedance of the structure as measured from the slit of the angular port that covers 30$^\circ$ of the azimuthal extent of the slit. These simulations  therefore provide $L_\text{eff}$. $L_\text{eff}$ and $|V|$ are then used in Equation \ref{eq:scanrate} to extract the scan rate. The ``external current density" function within COMSOL that models the physics of BES has been vetted and studied extensively for its accuracy. Appendix B of \cite{DMRadio:2023igr} discusses COMSOL techniques in more detail.

The noise physics encoded in $\bar{\mathcal{G}}[\omega_r,T,\eta(\omega_r)]$ of Equation \ref{eq:scanrate} depends on the noise temperature of the system, $T$, and the parameter $\eta(\omega_r)$, which is the frequency dependent ratio of the minimum noise temperature of an amplifier to the quantum limit at the resonance frequency $\omega_r$:
\begin{equation}
    \eta(\omega_r)\equiv \frac{k_BT_\text{min}(\omega_r)}{\hbar\omega_r/2}.
\end{equation}
The optimal frequency-dependent values of $\eta(\omega_r)$ using limits set by dc SQUIDs optimized for operation in an experiment like DMRadio are presented in Figure 3 of ~\cite{DMRmSquids}. These values are used to calculate the scan rate. The system temperature is evaluated at 50 mK.

\begin{figure*}
    \centering
    \includegraphics[width=.985\textwidth]{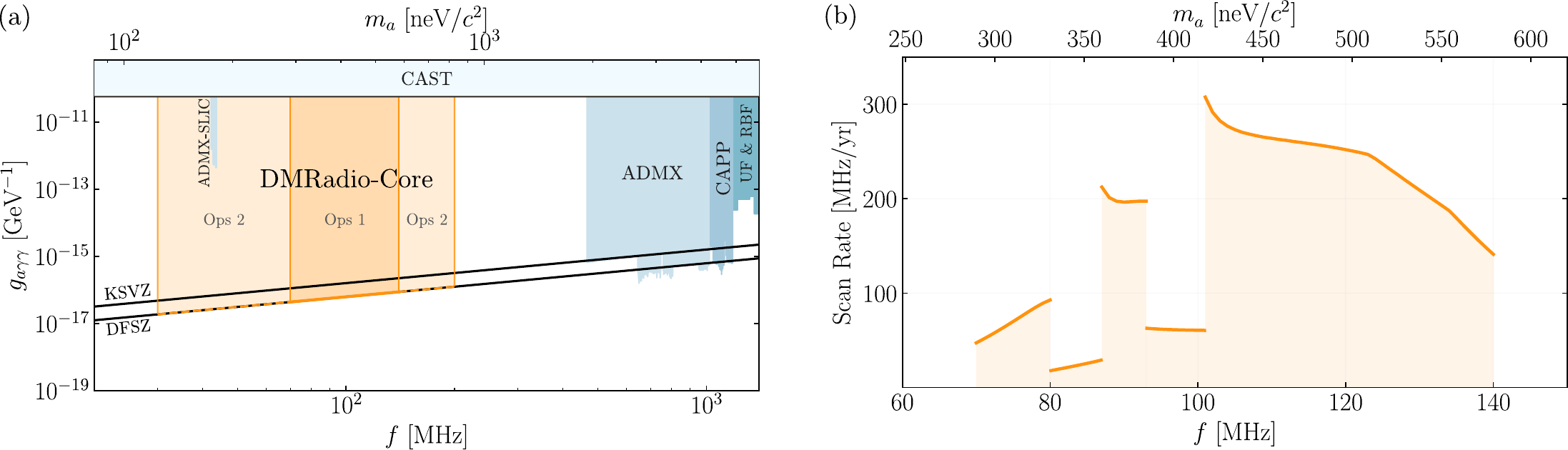}
    \caption{(a) DMRadio-Core is sensitive to DFSZ axions over the 70--140 MHz frequency range (290--580 neV/$c^2$) for Ops 1. The sensitivity assumes SNR=3 with a local dark matter density of $\rho_\text{DM}=0.45\text{ GeV cm}^{-3}$ and a quality factor of $Q=10^6$. The live scan time for this region is 0.75 years. Ops 2 of the experiment will search for DFSZ axions in the 30--200 MHz range. The CAST collaboration has the most competitive limit for $g_{a\gamma\gamma}$ in these frequencies at $g_{a\gamma\gamma}\lesssim5.7\times10^{-11}\text{ GeV}^{-1}$   \cite{CAST:2024eil} with the ADMX-SLIC experiment having constrained dark matter axions around 42 MHz at $g_{a\gamma\gamma}\lesssim4\times10^{-13}\text{ GeV}^{-1}$\cite{Crisosto:2019fcj}.
    (b) The scan rate of DMRadio-Core over the Ops 1 range of 70--140 MHz. Since each pickup size has different unfavorable frequency regions where the scan rate is low, some frequencies only utilize one size of pickup. Specifically the 80--86 MHz range only utilizes the $d_o=120$ cm pickup and the 94--100 MHz only utilizes the $d_o=146$ cm pickup. The data for the scan rates is available at \cite{corescanrates}.  
    }
    \label{fig:ScanRatesCORE}
\end{figure*}

The calculated scan rates for DMRadio-Core are shown in Figure \ref{fig:ScanRatesCORE}b over the range of  \mbox{70--140 MHz}. Since some pickup sizes can result in prohibitively slow scan rates or require unphysical tuning parameters, they are omitted from the axion search within those frequency ranges. This results in discontinuities in the scan rate. In addition, these scan rates are limited to a constant quality factor of $Q=10^6$ for all frequencies, as motivated by recent results of high $Q$ lumped superconducting resonators \cite{Kolevatov:2025lzx}.  For $\text{SNR}=3$, the scan rate corresponds to a live scan time of 0.75 years.

\section{The Core Geometry for DMRadio-GUT}
\label{sec:GUT}

DMRadio-GUT is the ultimate low-mass axion experiment within the DMRadio series with sensitivity to axions in the 100 kHz--30 MHz frequency range, corresponding to 0.4--120 neV$/c^2$ masses at DFSZ sensitivity, as shown in Figure \ref{fig:GUTScanRate}. To achieve the required sensitivity to such axions, this experiment must employ quantum sensors operating beyond the standard quantum limit using backaction evading techniques, high magnetic fields over large volumes, and high quality factors \cite{DMRadio:2022jfv}. While the target values for each of these parameters can be estimated using scaling arguments from experiments like DMRadio-m$^3$ and DMRadio-50L, solenoidal and toroidal pickup geometries each pose separate challenges for DMRadio-GUT in terms of achieving high magnetic fields, large volumes, and high quality factors. The benefits of the Core geometry as described in the previous section can mitigate these challenges. Here we provide a preliminary design of a Core magnet and pickup geometry as part of an experiment that achieves the target sensitivity of DFSZ axions at masses down to 0.4 neV/$c^2$.

\subsection{Magnet and pickup design}

\begin{figure}[ht]
    \centering
    \includegraphics[width=.995\columnwidth]{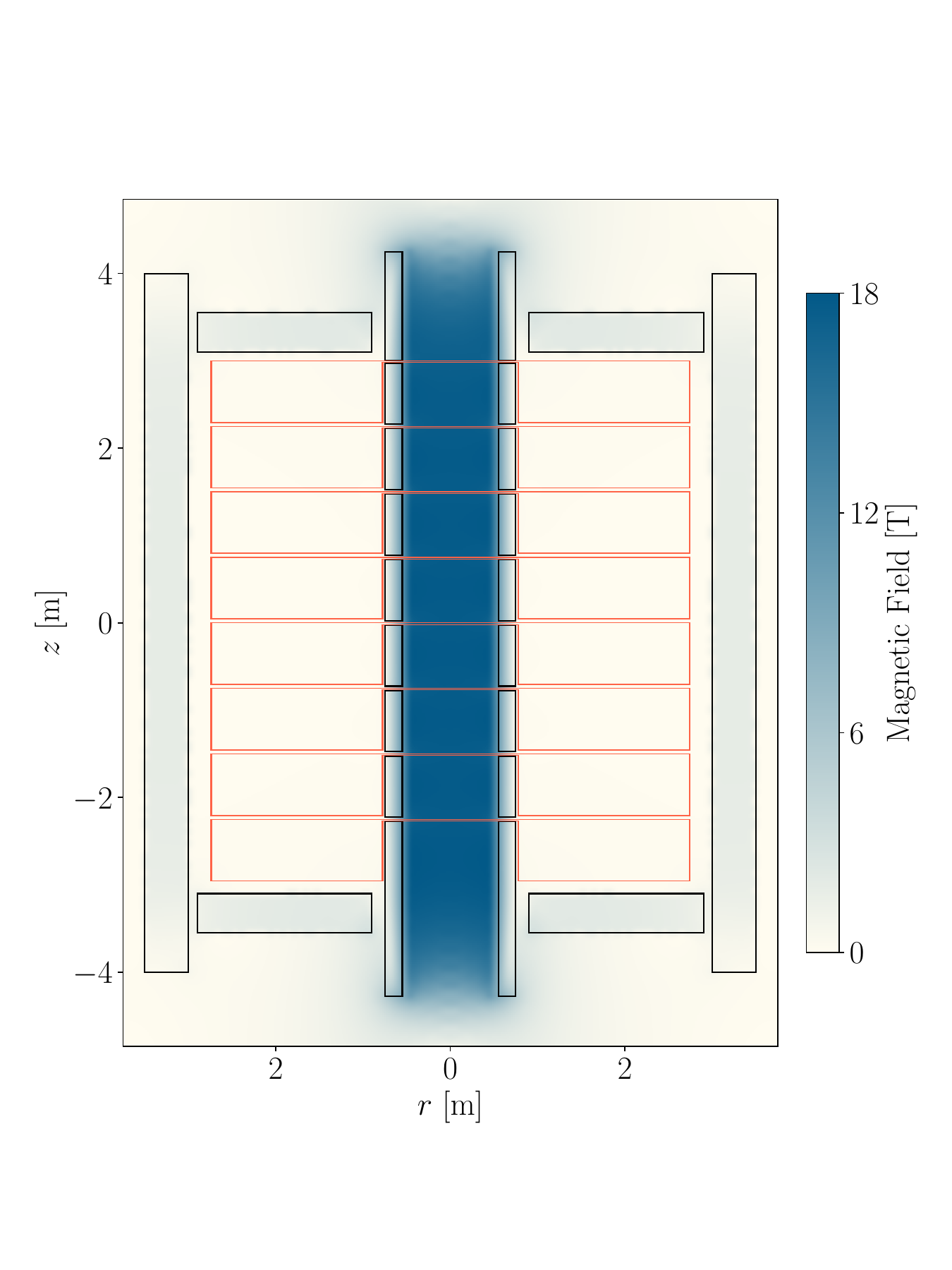} 
    \caption{Motivated by the magnet design presented in the previous section, an experimental setup that achieves the goals of DMRadio-GUT employs an 18 T peak field segmented solenoidal magnet with a 90 cm bore. The pickups (red) traverse the high-field region and each have a radius of \mbox{2.9 m}. Soft iron is also included in this design to reduce the dc magnetic field within the pickup region. 
    }
    \label{fig:GUT}
\end{figure}

\begin{figure}[ht]
    \centering
    \includegraphics[width=.995\columnwidth ]{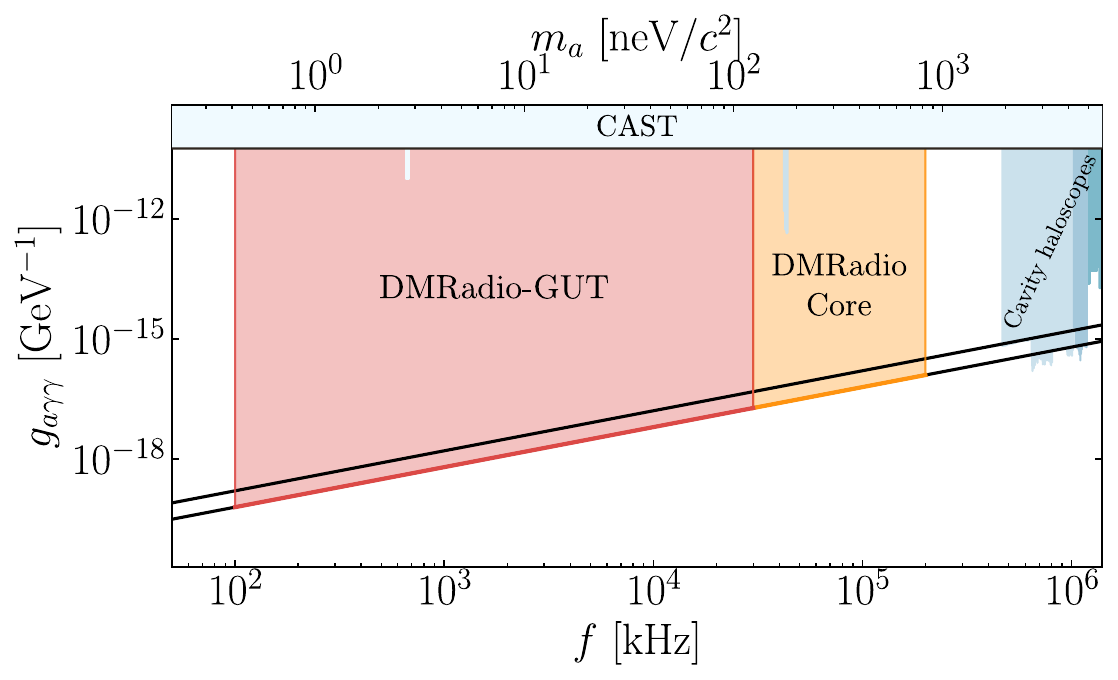}
    \caption{The sensitivity of the DMRadio-GUT experiment alongside DMRadio-Core. DMRadio-GUT is sensitive to axions at DFSZ sensitivity in the 100 kHz--30 MHz range using a 90 cm bore, 8.5 m long, 18 T magnet to achieve a scan time of 2.2 years. DMRadio-Core relies on a 37 cm bore, 3 m long, 5 T magnet. 
    }
    \label{fig:GUTScanRate}
\end{figure}

A design for this experiment is shown in Figure \ref{fig:GUT}. The magnet is an 18 T peak field solenoidal magnet, 8.5 m tall with a 90 cm bore diameter. Its long narrow profile achieves the desired effect of forcing the field lines to return at sufficiently large radii such that a low-field region exists outside of the magnet coils. Additionally, the use of three room-temperature soft iron segments reduces the dc field to $B_\text{dc}<170$ mT in the region outside the coils such that the entire region is below the $H_\text{c1}$ of niobium.  The solenoid itself is segmented into a total of 9 coils such that the pickup crosses the high-field region.

The pickups for such an experiment are shown in Figure \ref{fig:GUT} in red, overlaid on the magnetic field profile. The pickups are 2.9 m in radius and 75 cm in height.

\subsection{Science reach and scan time}

The finite element techniques presented in the previous section are used to estimate the scan time. Using projections for future sensors and resonators,  the scan rate is calculated using $Q=2\times 10^7$ and $\eta=0.1$ which corresponds to 20 dB of backaction evasion \cite{DMRadio:2022jfv}. These values alongside a value of $\text{SNR}=3$ produce the scan times tabulated in Table \ref{tab:scantime} for frequencies covering \mbox{100 kHz--30 MHz}. The live scan time of this experiment is 2.2 years. Specifically, the scan time is heavily dominated by the lowest frequencies since the 100--200 kHz range consists of 95\% of the total scan time.

\begin{table}[]
    \centering
\begin{tabular}{c|c} Magnet Specs
 &\begin{tabular}[c]{@{}c@{}} Live Scan Time \\ 100 kHz--30 MHz
 \end{tabular} \\ \hline
 $1\times18$ T magnet  &  2.2 yr                                     \\ \hline
$5\times 9$ T magnets   & 1.4  yr                            \\ \hline
$3\times18$ T magnets       & 
0.24 yr\\                                                   
\end{tabular}
    \caption{
Live scan times for different magnet configurations to achieve SNR$=3$. The $5\times9$ T and $3\times 18$ T configuration scan times are scaled analytically using the $1\times18$ T results by assuming that the magnets are all within one axion coherence length of each other.  In every magnet configuration, 95\% of the scan time is allocated to the 100--200 kHz range. The listed times do not take into account operations time or time dedicated to rescans.}
    \label{tab:scantime}
\end{table}

Alternative configurations and their corresponding scan rates are also presented in Table \ref{tab:scantime}. A set of five 9 T magnets with identical coil packs lead to a $\sim$40\% decrease in scan time relative to a single 18 T magnet. A set of three 18 T magnets lead to a decrease in the scan time by a factor of 9. In all scenarios, we assume that the different magnets are operating within the de Broglie wavelength of the dark matter axions.

Magnet technologies that could support such designs have delivered promising results, as summarized in \cite{DMRadio:2022jfv}. Nb$_3$Sn magnets can produce fields as high as 16 T and have been used for 12 T magnets at ITER for volumes that will ultimately store at least an order of magnitude more magnetic field energy than what is required for this design \cite{4497941}. High-$T_c$ superconductors such as YBCO and REBCO have also shown promise by producing fields as high as 35 T \cite{6649987}.  

While the lowest order cavity resonances of this design appear near 10 MHz thus complicating the scan strategy at the higher end of the target frequencies, the scan rates at these frequencies are well above 1 GHz/yr. As such, a modified set of smaller pickups with different dimensions can easily scan the upper end of the frequency range. A design of such pickups is beyond the scope of this paper.

\section{Conclusion}

High-field large-volume magnets pose one of the largest costs and technical challenges of scaling existing experiments to achieve the sensitivity required to search for axions at the neV$/c^2$ mass scale. In this paper, we have presented the Core geometry. The reduced magnet size required in the Core geometry provides a promising and appealing path towards the design and realization of such experiments thanks to a more efficient coupling between a pickup structure and the axion-induced electromagnetic response from the magnet.

Using this geometry we have presented the design of the DMRadio-Core experiment. This experiment utilizes a 5 T, 37 cm inner bore, 2.95 m tall segmented solenoidal magnet and is sensitive to DFSZ axions in the 30--200 MHz frequency range corresponding to masses of 120--830 neV/$c^2$. In addition to its science goals covering unexplored parameter space, this experiment also acts as a pathfinder towards a large-scale realization of an experiment which has sensitivity to DFSZ axions in the 0.4--120 neV$/c^2$ (100 kHz--30 MHz) range. This provides  the first representative design parameters of such an experiment.

\begin{acknowledgments}

Stanford University and UC Berkeley gratefully acknowledge support from the Gordon and Betty Moore Foundation, grant numbers 7941 and 61988545-138055 respectively.
This work was supported by the U.S.
Department of Energy, Office of High Energy Physics
program under the QuantISED program, FWP 101260, and by the National Science Foundation under award 2411650.  Chelsea Bartram and Andrew Yi were supported by the Department of Energy, Laboratory Directed Research and Development program at SLAC National Accelerator Laboratory, under contract DE-AC02-76SF00515 and as part of the Panofsky Fellowship awarded to Chelsea Bartram. J. T. Fry is supported by the National Science Foundation Graduate Research Fellowship under Grant No. 2141064.

\end{acknowledgments}


\appendix
\section{Lumped element--cavity formalism correspondence in DMRadio-Core}

\label{sec:Cavity_v_LumpedElement}

To understand the calculation of the signal in lumped-element haloscopes such as DMRadio, and how it contrasts with cavity haloscopes, it is important to distinguish between the cavity regime and the magneto-quasistatic (lumped-element) regime at the level of the underlying Maxwell formalism.  In general, excitation of a resonant electromagnetic mode by a source current density $\mathbf{J}_{\mathrm{drive}}$ scales as
\begin{equation}
\left|
\int_V
\mathbf{E}_{\text{mode}}(\mathbf{x})
\cdot
\mathbf{J}_{\mathrm{drive}}(\mathbf{x})
\, d^3x
\right|^2.
\label{eq:general_overlap}
\end{equation}
In axion haloscopes the effective source current density is
$\mathbf{J}_\text{eff}$ as given in Eq. \ref{eq:Jeff},
so Eq.~(\ref{eq:general_overlap}) reduces to the familiar cavity form factor~\cite{Choi:2020wyr} proportional to
\begin{equation}
\left|
\int_V
\mathbf{E}_{\text{mode}}
\cdot
\mathbf{B}_{\mathrm{dc}}
\, dV
\right|^2,
\end{equation}
which quantifies the spatial overlap between a cavity eigenmode and the magnetized volume.  In the cavity regime the electromagnetic energy remains confined to a geometric eigenmode whose field profile is fixed by boundary conditions, and sensitivity is directly determined by this overlap.

In contrast, in the magneto-quasistatic limit relevant for DMRadio, the system is more naturally described in circuit language.  The axion-induced effective current density $\mathbf{J}_\text{eff}$ within the magnetized region generates a time-varying magnetic flux, which drives a global current mode of an LC resonator.  Once excited, this current distribution is determined by the impedance of the resonant circuit and is not confined to the region where $\mathbf{B}_{\mathrm{dc}}\neq 0$.  The pickup loop measures the magnetic flux produced by this global current mode, which may be maximal in the region where the dc magnetic field is negligible, in direct analogy with a transformer or current clamp.

Power transfer remains consistent with Maxwell’s equations and the mode overlap formalism.  From Poynting’s theorem, the time-average power delivered to the resonator is
\begin{equation}
\langle P \rangle
=
\frac{1}{2}
\text{Re}
\left[
\int_V
\mathbf{J}_\text{eff}
\cdot
\mathbf{E}^*
\, d^3x
\right],
\end{equation}
where $\mathbf{E}$ includes the electric field produced by induced currents in the pickup circuit.  Thus the resonator extracts energy from the axion field via the electromagnetic backreaction field it generates, exactly as in a current transformer.

However, it is sufficient to compute the voltage applied to the lumped-element circuit. The above volume integral does not need to be computed explicitly.  Once the voltage on the lumped-element equivalent circuit is known, the power can be found exactly using the circuit complex impedance. Furthermore, in the magneto-quasistatic regime, the voltage in the lumped-element circuit follows directly from circuit relations
\begin{equation}
V(t) = -\frac{d\Phi(t)}{dt}.
\end{equation}
Nonetheless, as discussed in Section \ref{sec:sciencereachformalism}, the formalisms used in this paper do not rely on this equation since the MQS assumptions are only valid at sufficiently low frequencies.

Thus, the backreaction mode-overlap formulation and the circuit description are equivalent statements of the same underlying Maxwell dynamics.  The pickup can reside in a region of near-zero static magnetic field, and still be sensitive. This distinction underlies the Core geometry.


\bibliographystyle{apsrev4-2}
\bibliography{dmradio_bibliography}


\end{document}